\documentclass[sigconf]{acmart}

\usepackage{hyperref}
\usepackage{url}
\usepackage{booktabs}
\usepackage{amsfonts}
\usepackage{nicefrac}
\usepackage{microtype}
\usepackage{makecell}
\usepackage{mathtools}
\usepackage{graphicx}
\usepackage{multirow}
\usepackage{enumitem}
\usepackage{color}
\usepackage{xcolor}
\usepackage{colortbl}
\usepackage{pifont}
\usepackage{ulem}
\usepackage{bbding}
\usepackage{wrapfig}
\usepackage{caption}
\usepackage{subcaption}

\usepackage{algorithm}
\usepackage{algorithmic}

\AtBeginDocument{%
  }

\setcopyright{acmlicensed}
\copyrightyear{2018}
\acmYear{2018}
\acmDOI{XXXXXXX.XXXXXXX}
\acmConference[Conference acronym 'XX]{Make sure to enter the correct
  conference title from your rights confirmation email}{June 03--05,
  2018}{Woodstock, NY}
\acmISBN{978-1-4503-XXXX-X/2018/06}

\begin{document}

\title{SCASRec: A Self-Correcting and Auto-Stopping Model for Generative Route List Recommendation}

\author{Chao Chen}
\authornote{Both authors contributed equally to this research.}
\email{cc201598@alibaba-inc.com}
\author{Longfei Xu}
\authornotemark[1]
\email{longfei.xl@alibaba-inc.com}
\affiliation{\institution{AMAP, Alibaba Group}
  \state{Beijing}
  \country{China}}

\author{Daohan Su}
\email{dhsu@bit.edu.cn}
\affiliation{%
  \institution{Beijing Institute of Technology}
  \state{Beijing}
  \country{China}}

\author{Tengfei Liu}
\email{12332470@mail.sustech.edu.cn}
\affiliation{%
  \institution{Southern University of Science and Technology}
  \state{Beijing}
  \country{China}}

\author{Hanyu Guo}
\email{guohanyu.ghy@alibaba-inc.com}
\affiliation{\institution{AMAP, Alibaba Group}
  \state{Beijing}
  \country{China}}

\author{Yihai Duan}
\email{duanyihai.dyh@alibaba-inc.com}
\affiliation{\institution{AMAP, Alibaba Group}
  \state{Beijing}
  \country{China}}

\author{Kaikui Liu}
\email{damon@alibaba-inc.com}
\author{Xiangxiang Chu}
\email{cxxgtxy@gmail.com}
\affiliation{\institution{AMAP, Alibaba Group}
  \state{Beijing}
  \country{China}}

\renewcommand{\shortauthors}{Chao Chen et al.}

\begin{abstract}
Route recommendation systems commonly adopt a multi-stage pipeline involving fine-ranking and re-ranking to produce high-quality ordered recommendations. However, this paradigm faces three critical limitations. First, there is a misalignment between offline training objectives and online metrics. Offline gains do not necessarily translate to online improvements. Actual performance must be validated through A/B testing, which may potentially compromise the user experience. Second, redundancy elimination relies on rigid, handcrafted rules that lack adaptability to the high variance in user intent and the unstructured complexity of real-world scenarios. Third, the strict separation between fine-ranking and re-ranking stages leads to sub-optimal performance. Since each module is optimized in isolation, the fine-ranking stage remains oblivious to the list-level objectives (e.g., diversity) targeted by the re-ranker, thereby preventing the system from achieving a jointly optimized global optimum. To overcome these intertwined challenges, we propose \textbf{SCASRec} (\textbf{S}elf-\textbf{C}orrecting and \textbf{A}uto-\textbf{S}topping \textbf{Rec}ommendation), a unified generative framework that integrates ranking and redundancy elimination into a single end-to-end process. SCASRec introduces a stepwise corrective reward (SCR) to guide list-wise refinement by focusing on hard samples, and employs a learnable End-of-Recommendation (EOR) token to terminate generation adaptively when no further improvement is expected. Experiments on two large-scale, open-sourced route recommendation datasets demonstrate that SCASRec establishes an SOTA in offline and online settings. SCASRec has been fully deployed in a real-world navigation app, demonstrating its effectiveness.
\end{abstract}




\maketitle

\section{Introduction}
\label{sec:intro}
Modern route recommendation systems in navigation universally adopt a multi-stage paradigm comprising recall, rough-ranking, fine-ranking, and re-ranking, which has become the standard practice in large-scale industrial applications~\cite{covington2016deep, zhou2018deep}.
In this context, the paradigm operates by first recalling a set of candidate routes upon receiving an origin–destination query, typically using classical pathfinding algorithms~\cite{A2005star,abraham2013alternative,delling2017customizable}, followed by a rough-ranking stage.
The process then proceeds to the ubiquitous two-stage ranking pipeline, where a fine-ranking stage~\cite{zhou2018deep, chang2023twin} estimates the relevance of individual routes, and a re-ranking stage~\cite{carbonell1998use, chen2018fast} refines the final ordered route list by modeling contextual interactions among them.
This workflow has been widely adopted to generate high-quality, unique route lists (detailed discussions in Appendix~\ref{app:rwk}).

\begin{figure}[t]
\centering
\includegraphics[width=0.47\textwidth]{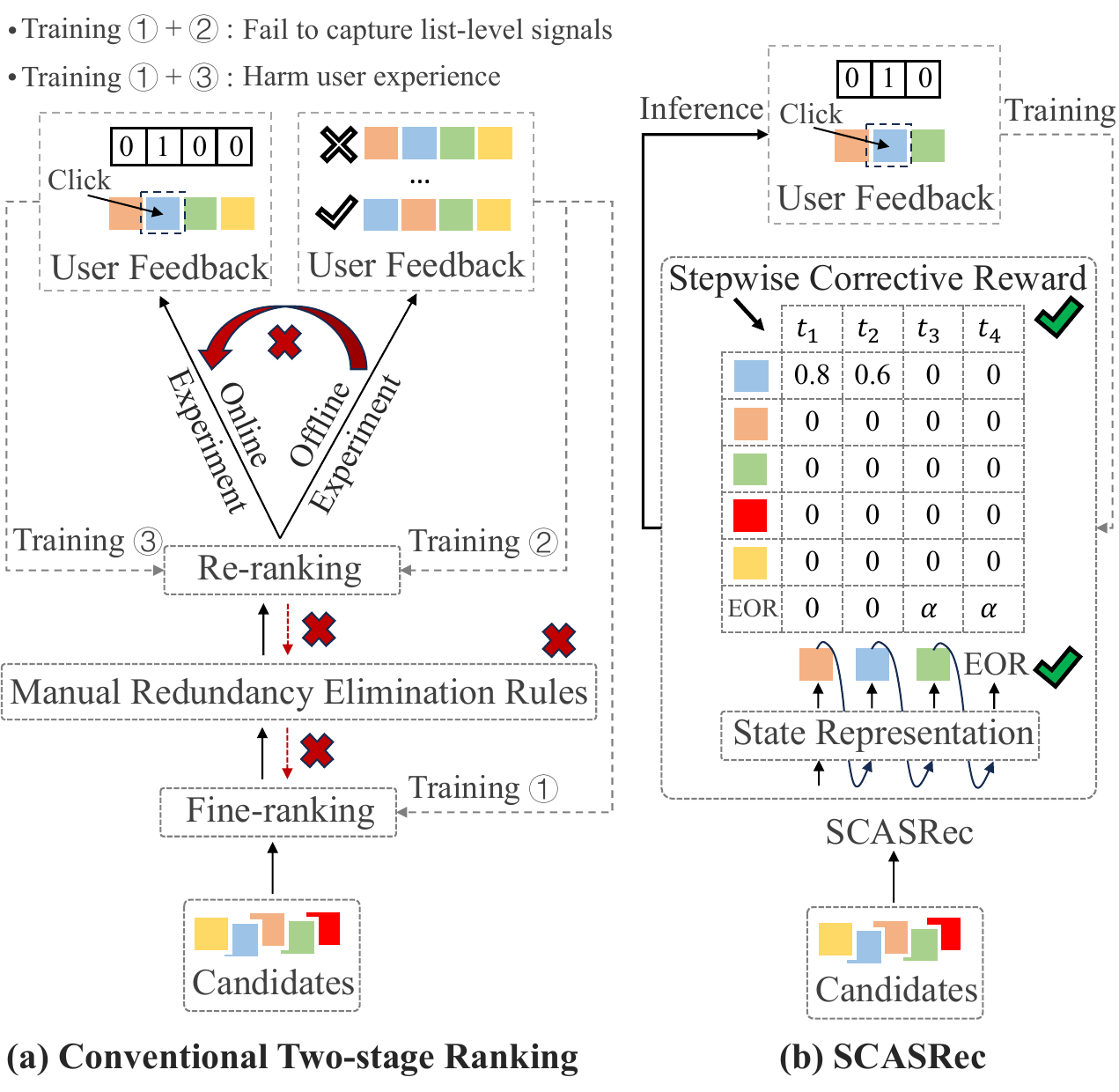}
\vspace{-0.3cm}
\caption{Comparison of two-stage ranking and SCASRec.}
\vspace{-0.6cm}
\label{fig:intro}
\end{figure}

Nonetheless, the conventional two-stage ranking paradigm encounters three fundamental limitations in list-level route recommendation, as depicted in Fig.~\ref{fig:intro}(a).
\textbf{Limitation~\ding{182}: Misalignment between offline objectives and online metrics.}
In practice, ranking models are typically trained on item-level signals (e.g., clicks), which correlate poorly with actual user satisfaction measured by list-level online metrics (e.g., coverage or diversity).
Consequently, improvements in offline loss often fail to translate into meaningful gains in user engagement, necessitating extensive A/B testing for validation.
This process is not only costly but may also degrade user experience during experimentation.
This disconnect between training objectives and real-world utility fundamentally limits the effectiveness of conventional pipelines.
To bridge this gap, it is essential to develop a reward that mirrors user intent and remains accessible via offline logs, independent of online intervention.

\textbf{Limitation~\ding{183}: Reliance on rigid manual redundancy rules.}
Current route recommendation systems typically eliminate redundancy through static, handcrafted heuristics rather than adaptive learning mechanisms.
These rules commonly enforce fixed thresholds based on predefined measures (e.g., discarding routes with similar ETA or a major detour) to filter out redundant routes and control list length.
However, such policies lack contextual awareness and fail to adapt to the high variance in user intent across different scenarios or domains.
For instance, the trade-off between speed and distance varies in urgent scenarios, just as price sensitivity differs among affluent users.
It is impractical to apply a static, universal filtering logic.
More critically, these rules optimize for objectives such as pairwise diversity, which are fundamentally misaligned with end-to-end recommendation metrics like MRR or coverage ratio.
This inspires a learnable approach to adaptively manage route termination, tailored to specific user utility.

\textbf{Limitation~\ding{184}: Iterative coupling between disjoint ranking stages.}
The strict separation between fine-ranking and re-ranking results in a fragmented pipeline.
The re-ranking module processes only the fixed output of fine-ranking and cannot provide feedback to refine its initial scoring, thereby precluding end-to-end joint optimization and trapping the system in a local optimum.
Such architectural decoupling not only propagates and amplifies errors across stages but also complicates system maintenance and iteration in real-world deployments.
This underscores the need for a unified framework that integrates candidate selection, contextual refinement, and redundancy control into a single process.


To address these interwined challenges, we propose \textbf{SCASRec}, a \textbf{S}elf-\textbf{C}orrecting and \textbf{A}uto-\textbf{S}topping model for generative route list \textbf{Rec}ommendation.
As illustrated in Fig.~\ref{fig:intro}(b), SCASRec unifies fine-ranking, re-ranking, and redundancy elimination into a single encoder-decoder architecture, generating route lists step by step and terminating adaptively.
The first key component is the \textbf{Stepwise Corrective Reward (SCR)}, introduced explicitly to bridge the gap between offline training and online user satisfaction.
Instead of relying solely on sparse item-level clicks, SCR leverages list-level signals (i.e., the List Coverage Rate (LCR) derived from user interactions) as an additional sequence-level supervision signal.
At each step, SCR evaluates the expected marginal gain of refining the current partial list toward better coverage of the ground-truth.
This stepwise feedback, combined with click labels, guides the model toward contextual, incremental corrections that directly optimize for online-aligned objectives, rather than isolated item relevance.
Second, SCASRec introduces an learnable \textbf{End-of-Recommendation (EOR)} token as a adaptive stopping criterion, replacing rigid handcrafted rules with a data-driven mechanism.
During training, the model is supervised to predict EOR immediately after the ground-truth route is generated.
At inference time, the recommendation process ends when the model generates the EOR token, enabling adaptive list lengths that dynamically respond to user intent and scenario context.
To further enhance robustness, we employ a heuristic noise-aware training strategy that adjusts the EOR reward based on estimated data quality.
By integrating SCR and EOR into a unified generative process, SCASRec overcomes the fragmented optimization of conventional two-stage pipelines.
The entire system is trained end-to-end with awareness of both ranking quality and list-level utility, allowing it to converge toward a globally coherent solution.
Together, these mechanisms empower SCASRec to generate concise, diverse, and high-quality route recommendation lists that align closely with actual user behavior.
To facilitate research in route recommendation, we release a large-scale route recommendation dataset comprising approximately 500,000 queries and 6 million candidate routes.
It includes rich features such as route attributes, user historical interactions, and road network topology, making it the most comprehensive public dataset available for route recommendation to date.


In summary, the main contributions of our work are as follows:
\ding{192} \textit{\underline{Unified Framework.}}
We propose SCASRec, a self-correcting and auto-stopping generative recommendation model that unifies fine-ranking, re-ranking, and redundancy elimination in a single pipeline, eliminating iterative coupling and manual post-processing.
\ding{193} \textit{\underline{Novel Mechanisms.}}
We introduce the SCR, a list-level supervision signal derived from offline interactions that directly aligns offline training with online metrics, and an EOR token with noise-aware training to dynamically terminate recommendations, replacing rigid redundancy rules.
\ding{194} \textit{\underline{SOTA Performance.}}
SCASRec achieves SOTA performance on both offline and online experiments and has been fully deployed in an online navigation application.
We also release a large-scale dataset with rich features to support future research.

\section{Preliminary}
\label{sec:preliminary}

\subsection{Notation and Problem Definition}
In the route recommendation task, a system receives an origin-destination query from a user and returns an ordered list of candidate routes. 
Formally, after the recall (route planning) and rough-ranking stages, we obtain a set of $N$ candidate routes, denoted as $\mathcal{P} = \left\{p_1, \dots, p_N\right\}$.
The goal of the ranking model is to generate a ranked list $\bar{P} = \left\{\bar{p}_1,\dots,\bar{p}_K\right\}$ with $K\leq N,\; \bar{p_i}\in\mathcal{P}$ that best matches the user's true preference.
The user's actual trajectory $u$ serves as implicit feedback to evaluate the quality of $\bar{P}$.

An ideal route recommendation system should simultaneously achieve three objectives:
(1) rank the user's preferred (ground-truth) route as high as possible,
(2) ensure high overall quality of the presented list, and
(3) avoid showing redundant routes after the preferred one has been found.
To this end, we define our optimization goal as maximizing a combined metric of ranking performance and list coverage, while minimizing redundant exposure.
A detailed formulation of these objectives, including the definitions of Mean Reciprocal Rank (MRR), List Coverage Rate (LCR), and the redundant item set $Z$, is provided in Appendix~\ref{app:problem_formulation}.

\subsection{Route Recommendation}
Traditional route planning methods rely on algorithms like A*~\cite{A2005star} and Dijkstra to find the shortest path.
To enhance diversity, subsequent work explored route penalization~\cite{route2013planning}, Pareto optimization~\cite{preferred2017path}, and multi-objective optimization~\cite{personalized2015route}.
In large-scale road networks, computational efficiency becomes critical, with techniques like Highway Hierarchies~\cite{Highway2005hierarchies} and parallel computing significantly reducing processing time.

However, route planning models are limited by their inability to provide a complete view of routes and are constrained by efficiency requirements, making complex models impractical in high-concurrency scenarios.
Thus, the industry currently treats route planning as a recall stage to generate a route set, followed by the route recommendation task.
Recent advances include ID-based embeddings, such as edge-level embeddings~\cite{cheng2021r4}, and multi-scenario models like DSFNet~\cite{yu2025dsfnet}, which outperform MMOE~\cite{ma2018modeling}.

Despite significant progress, these approaches still follow the multi-stage paradigm.
They primarily focus on item-level relevance scoring and rely on manually-defined rules for redundancy elimination.
Crucially, they lack a principled mechanism for list-level optimization, which requires understanding the contextual interactions between items in the list and making sequential decisions about both content and length.
This gap in the literature motivates our work.

\section{Method}
\label{sec:method}

\subsection{Model Overview}
\begin{figure*}[t]
\centering
\includegraphics[width=\textwidth]{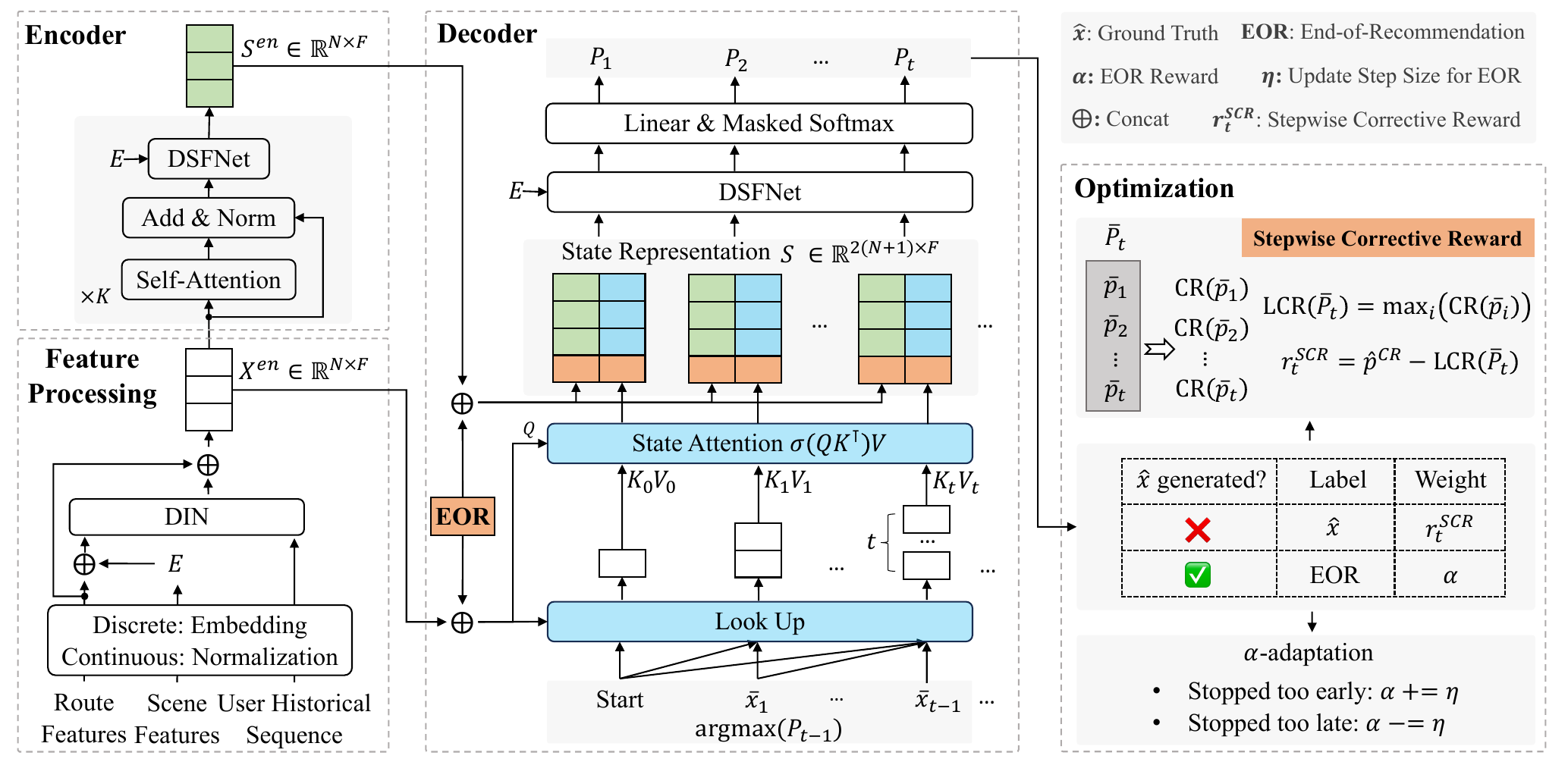}
\vspace{-0.8cm}
\caption{The generative framework of SCASRec for route list recommendation. SCR provides stepwise list-level feedback to guide sequential refinement, while the EOR token enables adaptive termination for redundancy control.}
\vspace{-0.2cm}
\label{fig:framework}
\end{figure*}

We propose SCASRec, a unified encoder-decoder framework that jointly optimizes ranking quality and redundancy control in an end-to-end manner.
As illustrated in Fig.~\ref{fig:framework}, SCASRec processes route features, scene context, and user historical interactions through a feature processing module, then encodes global item interactions via a multi-scenario self-attention mechanism.
The decoder sequentially generates the recommendation list by attending to previously selected routes and updating a stepwise state representation.
Detailed descriptions are provided in Appendix~\ref{app:detailed_model}.

Crucially, SCASRec targets a global objective that explicitly balances coverage and redundant exposure:
\begin{equation}
\label{eq:overall_objective}
\max_{\theta}\left(\text{MRR}(D)+\text{LCR}(D)-\alpha|Z|\right),
\end{equation}
where $\alpha>0$ controls the trade-off between coverage and conciseness.
To align sequential decoding with this global objective, SCASRec introduces two core mechanisms:
(i) the \textbf{Stepwise Corrective Reward (SCR)}, which provides list-aware feedback at each step to guide contextual refinement; and
(ii) the \textbf{End-of-Recommendation (EOR)} token, which serves as a learnable stopping criterion to eliminate redundancy without manual rules.
The following subsections detail how SCR and EOR jointly optimize Eq.~\eqref{eq:overall_objective}.



\subsection{Stepwise Corrective Reward}
\label{sec:scr}
Conventional ranking models are typically trained on sparse item-level signals (e.g., clicks).
However, this offline objective is fundamentally misaligned with online user satisfaction, which is better reflected by list-level metrics like trajectory coverage, as discussed in \textbf{Limitation \ding{182}} of Sec.~\ref{sec:intro}.
To bridge this gap, we introduce the SCR, a list-wise signal derived from offline user trajectories that directly aligns the training process with online-aligned objectives.

Formally, let $\bar{P}_t$ denote the recommendation route list generated up to step $t$, and let $\hat{p}^{\text{CR}}$ be the coverage rate of the ground-truth route.
We define the SCR at step $t$ as:
\begin{equation}
r_t^{\text{SCR}}=\hat{p}^{\text{CR}}-\text{LCR}\left(\bar{P}_t\right),
\label{eq:scr}
\end{equation}
which represents the remaining gap between the current list coverage and the optimal coverage.
A larger $r_t^{\text{SCR}}$ indicates greater potential gain from additional corrections, signaling that the sample requires more attention during training.

\begin{figure}[t]
\centering
\includegraphics[width=0.48\textwidth]{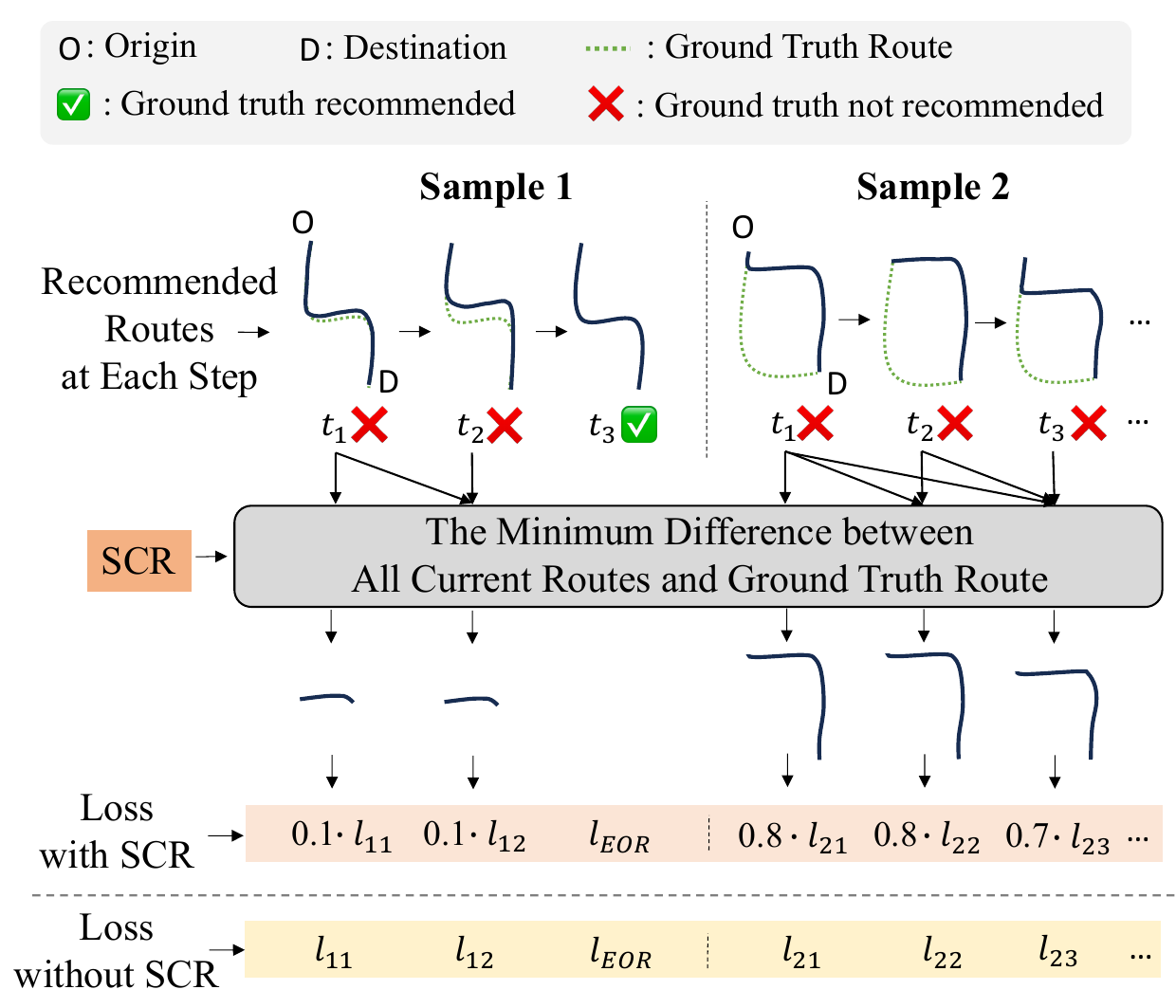}
\vspace{-0.8cm}
\caption{The SCR mechanism in route recommendation.}
\vspace{-0.5cm}
\label{fig:scr}
\end{figure}

As shown in Fig.~\ref{fig:scr}, SCR dynamically reflects the room for improvement at each step by measuring the minimal coverage gap between the current list and the ground-truth route.
This focused signal steers training loss toward steps with the highest potential gain, enabling SCASRec to rapidly converge toward high-quality, non-redundant lists that closely match user intent.

This design directly aligns with the coverage term in our global objective in Eq.~\eqref{eq:overall_objective}.
By prioritizing samples with high $r_t^{\text{SCR}}$, the model accelerates the inclusion of the ground-truth route in the top positions, thereby simultaneously improving both LCR and MRR.
Once the ground-truth route is included, $\text{LCR}\left(\bar{P}_t\right)$ reaches $\hat{p}^{\text{CR}}$, causing $r_t^{\text{SCR}}$ to drop to zero and signaling that further additions provide negligible gain in either coverage or ranking quality.

Moreover, because routes similar to those already recommended contribute little to increasing $\text{LCR}\left(\bar{P}_t\right)$, they result in only marginal reductions in $r_t^{\text{SCR}}$.
In contrast, diverse alternatives that significantly expand trajectory coverage lead to larger reward drops, implicitly encouraging the model to explore meaningfully distinct options.
This mechanism promotes recommendation diversity without explicit constraints or post-hoc filtering.





\subsection{End-of-Recommendation}
\label{sec:eor}
Current route recommendation systems rely on rigid, handcrafted rules to eliminate redundancy, which is a practice that lacks adaptability across diverse user intents and scenarios, as highlighted in \textbf{Limitation~\ding{183}} of Sec.~\ref{sec:intro}.
To replace these heuristics with a learnable stopping mechanism, we introduce the EOR token, which explicitly optimizes the redundancy term $|Z|$ in Eq.~\eqref{eq:overall_objective}.

Specifically, let $\hat{t}$ denote the step at which the ground-truth route is first included in the generated list.
Since any route recommended after $\hat{t}$ contributes to $|Z|$, the optimal policy should terminate immediately at step $\hat{t}+1$.
We therefore assign a positive reward $\alpha>0$ to select EOR at step $\hat{t}+1$, and zero reward otherwise:
\begin{equation}
r_t^{\text{EOR}} =
\begin{cases}
\alpha, & \text{if } t=\hat{t}+1, \\
0,      & \text{otherwise}.
\end{cases}
\label{eq:eor}
\end{equation}
This reward makes the EOR a direct signal for the redundancy penalty $-\alpha |Z|$, enabling the model to learn not only \textit{what} to recommend but also \textit{when} to stop.

The trade-off coefficient $\alpha$ controls the aggressiveness of early termination.
Rather than fixing $\alpha$ manually, we employ a lightweight noise-aware adaptation strategy that dynamically adjusts $\alpha$ during training based on an estimated noise ratio $\beta$.
This allows the stopping behavior to automatically align with data quality and business requirements without extensive hyperparameter tuning.
Full details of the update rule are provided in Appendix~\ref{appendix:alpha}.

\subsection{Optimization and Training}
\label{sec:optimization}
Conventional two-stage route ranking pipelines suffer from fragmented optimization, mentioned in \textbf{Limitation~\ding{184}}.
In contrast, SCASRec enables end-to-end learning by unifying ranking, refinement, and adaptive stopping into a single generative process trained with supervised signals.

Formally, at each decoding step $t$, the model outputs a probability distribution $P_t\in\mathbb{R}^{N+1}$ over the $N$ candidate routes and the EOR token.
Let $\hat{t}$ denote the step at which the ground-truth route is first recommended.
The ground-truth label $Y_t$ is then defined as:
\begin{equation}
Y_t[i] =
\begin{cases}
1, & \text{if } t\leq\hat{t} \text{ and } i=\text{index}(\hat{p}), \\
1, & \text{if } t=\hat{t}+1 \text{ and } i=\text{index}(\text{EOR}), \\
0, & \text{otherwise},
\end{cases}
\label{eq:label}
\end{equation}
where $\text{index}(\cdot)$ maps an item to its position in the candidate set.
No loss is computed for steps beyond $t=\hat{t}+1$.

To incorporate list-level feedback, we weight the supervised loss at each step using the combined reward:
\begin{equation}
r_t=r_t^{\text{SCR}}+r_t^{\text{EOR}}.
\label{eq:r}
\end{equation}
The training objective is a weighted cross-entropy loss:
\begin{equation}
\mathcal{L}=-\sum_{t=1}^{\hat{t}+1}r_t\cdot Y_t\cdot\log (P_t).
\label{eq:loss}
\end{equation}
The overall training process is shown in Algorithm~\ref{alg}.

Critically, our primary training paradigm is fully supervised.
Actions are always drawn from historical user behavior, and the rewards only modulate loss weights rather than determine action selection.
This avoids the high variance and unsafe exploration inherent in reinforcement learning (RL).
For completeness, we also describe an RL variant of SCASRec in Appendix~\ref{app:rl}, which serves as a comparative baseline in our experiments.

\subsection{Theoretical Analysis}
\label{sec:theory}
In this section, we establish a formal theoretical foundation for the superiority of the SCASRec framework over conventional two-stage ranking pipelines.
We demonstrate that the global list-wise objective defined in Eq.~\eqref{eq:overall_objective}, which directly reflects online user-centric metrics, admits a well-defined global optimum.
Crucially, the unified generative architecture of SCASRec is capable of recovering this optimum, whereas the structural constraints of a two-stage pipeline inherently limit it to sub-optimal local solutions.
We begin by formally defining the optimal policy with respect to our objective:
\begin{equation}
F\left(\bar{P}\right) = \text{MRR}\left(\bar{P}\right) + \text{LCR}\left(\bar{P}\right) - \alpha\left|Z\left(\bar{P}\right)\right|.
\end{equation}
Let $\hat p$ denote the ground-truth route for a given query, and let $\text{CR}\left(\hat p\right)$ be its coverage rate.
The following policy $\pi^*$ achieves the maximum possible value of $F$:
\begin{equation}
\pi^*: \bar p_1=\hat p, \; \bar p_2=\text{EOR},
\end{equation}
where $\bar p_t$ represents the action selected at decoding step $t$.
The resulting list $\bar{P}^*=\left\{\hat p\right\}$ yields $\text{MRR}(\bar{P}^*)=1$, $\text{LCR}(\bar{P}^*)=\text{CR}(\hat p)$, and $|Z(\bar{P}^*)|=0$, leading to the optimal objective value:
\begin{equation}
F(\bar{P}^*)=1+\text{CR}\left(\hat p\right).
\end{equation}
No other list can achieve a higher MRR or LCR, and any extension of $\bar{P}^*$ introduces redundant items ($\left|Z\right|>0$), thereby decreasing $F$.

The SCASRec is explicitly designed to guide the learning process towards this optimal policy $\pi^*$.
Its supervised training objective provides direct and unambiguous signals for both key actions of $\pi^*$.
First, the SCR creates persistent learning pressure to include $\hat p$ as early as possible, since $\text{LCR}\left(\bar{P}_t\right)<\text{CR}\left(\hat p\right)$ for any partial list $\bar{P}_t$ that does not contain $\hat p$, resulting in a positive reward weight that prioritizes its selection.
Second, the EOR is supervised with a ground-truth label immediately following the inclusion of $\hat p$, providing a direct signal for optimal termination that minimizes redundancy.
This end-to-end supervision ensures that the model's optimization landscape contains a clear path to the global optimum $\pi^*$.
In contrast, a conventional two-stage pipeline is structurally incapable of reliably recovering $\pi^*$.
A detailed discussion can be found in Appendix~\ref{app:theory}.

\begin{table*}[t]
\centering
\caption{Performances in the offline setting on our dataset. The best results are highlighted in \textbf{Bold}.}
\vspace{-0.3cm}
\resizebox{\textwidth}{!}{
\renewcommand\arraystretch{1.1}
\begin{tabular}{lccccccccccc}
\Xhline{1pt}
\rowcolor{gray!16} \textbf{Method} & \textbf{HR@1} & \textbf{HR@2} & \textbf{HR@3} & \textbf{HR@4} & \textbf{HR@5} & \textbf{LCR@1} & \textbf{LCR@2} & \textbf{LCR@3} & \textbf{LCR@4} & \textbf{LCR@5} & \textbf{MRR}\\
\hline
MMR & 62.53 & 79.09 & 86.41 & 90.48 & 93.14 & 78.51 & 86.68 & 90.17 & 91.96 & 93.06 & 0.478\\
DNN & 62.62 & 78.91 & 86.28 & 90.45 & 93.05 & 78.52 & 86.56 & 90.04 & 91.89 & 92.99 & 0.475 \\
DPP & 60.55 & 77.74 & 85.67 & 90.10  & 92.76 & 77.49 & 86.80 & 90.34 & 92.11  & 93.15 & 0.452\\
PRM & 70.38 & 84.38 & \textbf{90.26} & 93.61 & 95.49 & 82.76 & 89.55 & 92.26 & 93.47 & 94.15 & 0.548 \\
Seq2Slate & 63.35 & 79.67 & 87.09 & 91.01 & 93.55 & 79.35 & 87.73 & 90.96 & 92.56 & 93.58 & 0.490\\
NAR4Rec & 67.37 & 72.08 & 75.48 & 78.66 & 81.96 & 81.31 & 83.86 & 85.93 & 87.81 & 89.36 & 0.291\\
\hline
\rowcolor[HTML]{D7F3F9} SCASRec+RL & 68.57 & 82.83 & 88.74 & 92.07 & 94.17 & 82.06 & 88.60 & 91.17 & 92.58 & 93.42 & 0.536\\
\rowcolor[HTML]{D7F3F9} SCASRec & \textbf{71.56} & \textbf{87.78} & 89.92 & \textbf{95.19} & \textbf{96.98} & \textbf{82.84} & \textbf{91.48} & \textbf{92.54} & \textbf{94.52} & \textbf{94.96}  & \textbf{0.590} \\
\Xhline{1pt}
\end{tabular}}
\label{tab:off_data}
\end{table*}

\begin{table*}[t]
\centering
\caption{Performances in the offline setting on the MSDR dataset.
The best results are highlighted in \textbf{Bold}.}
\vspace{-0.3cm}
\resizebox{0.7\textwidth}{!}{
\renewcommand\arraystretch{1.1}
\begin{tabular}{lccccccc}
\Xhline{1pt}
\rowcolor{gray!16} \textbf{Method} & \textbf{HR@1} & \textbf{HR@2} & \textbf{HR@3} & \textbf{LCR@1} & \textbf{LCR@2} & \textbf{LCR@3} & \textbf{MRR} \\
\hline
MMR          & 37.31 & 70.38 & 94.10 & 58.81 & 76.44 & 84.90 & 0.501 \\
DNN          & 35.67 & 71.71 & 94.27 & 57.97 & 76.83 & 84.94 & 0.501 \\
DPP          & 38.34 & 70.41 & 94.69 & 59.03 & 76.49 & 85.09 & 0.508 \\
PRM          & 36.39 & 72.42 & 94.82 & 58.34 & 77.07 & 85.11 & 0.506 \\
Seq2Slate    & 36.85 & 73.37 & 94.31 & 58.51 & 77.65 & 85.03 & 0.511 \\
NAR4Rec      & \textbf{42.70} & 76.87 & 91.28 & 60.46 & 78.11 & 83.85 & 0.487 \\
\hline
\rowcolor[HTML]{D7F3F9} SCASRec+RL & 32.85 & 77.54 & 91.82 & 56.70 & 79.00 & 84.26 & 0.506 \\
\rowcolor[HTML]{D7F3F9} SCASRec & 42.64 & \textbf{77.65} & \textbf{94.92} & \textbf{61.11} & \textbf{79.21} & \textbf{85.23} & \textbf{0.541} \\
\Xhline{1pt}
\end{tabular}}
\label{tab:off_msdr}
\end{table*}

\begin{table}[t]
\centering
\caption{Examples of the key features provided in our dataset.}
\vspace{-0.3cm}
\resizebox{0.48\textwidth}{!}{
\renewcommand\arraystretch{1.1}
\begin{tabular}{l|c|l}
\Xhline{1pt}
Feature Type & Shape & Some Key Features \\
\hline \hline
\multirow{2}{*}{Route Features} & \multirow{2}{*}{$N\times 62$} & The estimated time of arrival for the route \\
& & The total distance length of the route \\
\hline
\multirow{2}{*}{Scene features} & \multirow{2}{*}{$1\times 10$} & Request time \\
& & User familiarity with the origin and destination \\
\hline
\multirow{2}{*}{User Historical Seq} & \multirow{2}{*}{$T\times31$} & Selected route features \\
& & Unselected route features \\
\Xhline{1pt}
\end{tabular}}
\label{tab:datafeat}
\end{table}

\section{Experiments}
We evaluate SCASRec on two large-scale real-world route recommendation datasets, including a new benchmark that will be publicly released and MSDR~\cite{yu2025dsfnet} (detailed discussion in Appendix~\ref{app: dataset}).
Our comparison includes representative baselines spanning diversity-based methods (MMR~\cite{carbonell1998use}, DPP~\cite{chen2018fast}), context-aware models (DNN~\cite{covington2016deep}, PRM~\cite{pei2019personalized}, Seq2Slate~\cite{bello2018seq2slate}), and a generative approach (NAR4Rec~\cite{ren2024non}).
Performance is assessed via both offline metrics (HR@K, LCR@K, MRR) and online A/B tests measuring user engagement and operational efficiency.

\subsection{Detailed Experiment Settings}
\subsubsection{Baselines}
We compare SCASRec against a diverse set of representative baselines, covering three major paradigms in list-wise recommendation:
(1) diversity-aware methods (MMR, DPP),
(2) context-aware ranking models (DNN, PRM, Seq2Slate), and
(3) generative list construction approaches (NAR4Rec).
Below, we briefly summarize each method.

\noindent \textbf{\textbullet\ MMR}~\cite{carbonell1998use}.
Maximal Marginal Relevance (MMR) is a diversification algorithm that iteratively selects items with high relevance to the query and low redundancy to previously selected items.

\noindent \textbf{\textbullet\ DNN}~\cite{covington2016deep}.
The deep neural network is a basic deep learning method for CTR prediction, utilizing MLP to capture high-order feature interactions.

\noindent \textbf{\textbullet\ DPP}~\cite{chen2018fast}.
Determinantal Point Processes is an algorithm based on the Determinantal Point Process, which maximizes subset probabilities through Fast Greedy Map Inference.

\noindent \textbf{\textbullet\ PRM}~\cite{pei2019personalized}.
The Personalized Re-ranking Model (PRM) employs the self-attention mechanism to capture the mutual influence between items in the recommendation list.

\noindent \textbf{\textbullet\ Seq2Slate}~\cite{bello2018seq2slate}.
Seq2Slate is a reranking method based on the sequence-to-sequence framework, leveraging RNN to directly generate the final ranking results.

\noindent \textbf{\textbullet\ NAR4Rec}~\cite{ren2024non}.
NAR4Rec uses a non-autoregressive generative model to generate the final ranking in parallel, efficiently capturing global dependencies in the sequence.

\subsubsection{Evaluation Metrics}
We validate SCASRec through both offline and online experiments.
Differences in user feedback mechanisms lead to slight variations in the evaluation metrics for each, as detailed below:

\textbf{Offline Experiments.}
We adopt list-wise metrics that reflect both ranking quality and user satisfaction:

\noindent \textbf{\textbullet\ HR@K} measures whether the user’s actual traveled route appears in the top-$K$ recommendations.

\noindent \textbf{\textbullet\ LCR@K} quantifies the coverage between the recommended routes and the ground-truth trajectory.

\noindent \textbf{\textbullet\ MRR} evaluates the ranking position of the optimal route.
When MRR is equal, a higher LCR reflects better recommendation performance.

\textbf{Online Experiments.}
In online A/B tests, we report HR@K and LCR@K together with several key operational metrics.

\noindent \textbf{\textbullet\ Routes} denotes the average number of routes presented to users per session.

\noindent \textbf{\textbullet\ Deviation Rate (DR)} is the fraction of navigation sessions in which users deviate from the recommended route.

\noindent \textbf{\textbullet\ Low Diversity Ratio (LDR)} measures the proportion of impressions where the recommended routes exhibit insufficient inter-route diversity.

\noindent \textbf{\textbullet\ Redundant Route Ratio (RRR)} quantifies the proportion of recommended routes that are judged by domain experts as redundant or unlikely to be selected by users.

Both LDR and RRR are assessed through manual evaluation on a sampled subset of experimental traffic data.

\subsubsection{Experimental Environments}
SCASRec is trained on 8 H20 GPUs with the batch size set to 128 and the learning rate set to 0.001 using the Adam~\cite{adam2014method} optimizer. The training process completes 300k steps within 24 hours.

\subsection{Offline Experiments}
To validate the effectiveness of SCASRec, we conduct comprehensive offline evaluations on our large-scale route recommendation dataset and the public MSDR dataset~\cite{yu2025dsfnet}.
For evaluation, we consider two training paradigms for SCASRec: the primary supervised learning and a reinforcement learning (RL) variant.
This allows us to assess the impact of the optimization strategy on performance.
All baselines employ the same feature processing as described in Appendix~\ref{app:feature_process} to ensure a fair comparison.

On our route recommendation dataset (Table~\ref{tab:off_data}), SCASRec consistently outperforms all baselines across all metrics.
Specifically, it achieves 71.56\% HR@1, significantly outperforming the second-best method (PRM at 70.38\%), which demonstrates its superior ability in placing the user’s actual chosen route at the top position.
The performance advantage is sustained and even amplified at higher ranks.
SCASRec attains 96.98\% HR@5, the highest among all models.
Similarly, on LCR@K, which measures the coverage between the recommended list and the ground truth, SCASRec leads by a clear margin.
This indicates that its recommendations are not only highly relevant but also better aligned with users’ true behavior from a list-level perspective.
The highest MRR further confirms that SCASRec consistently ranks relevant routes earlier than competitors.

We also observe significant and consistent gains on the public MSDR dataset, as shown in Table~\ref{tab:off_msdr}.
SCASRec achieves the best overall performance, securing the highest MRR (0.541) and leading across most metrics.
It attains 42.64\% HR@1, nearly matching the strongest baseline, while significantly outperforming baselines in HR@2 (77.65\%) and HR@3 (94.92\%), demonstrating its robustness in capturing user intent beyond the top position.
Most importantly, it significantly surpasses all competitors on list-level coverage metrics, achieving the highest LCR@K.
This confirms that our unified framework’s explicit optimization of trajectory coverage generalizes effectively to external datasets with distinct data distributions, validating the adaptability and strong generalization capability of SCASRec in diverse real-world route recommendation scenarios.

\begin{table*}[t]
\centering
\caption{Ablation study of SCASRec with and without SCR and EOR on our route recommendation dataset.}
\vspace{-0.3cm}
\resizebox{\textwidth}{!}{
\renewcommand\arraystretch{1.1}
\begin{tabular}{lccccccccccc}
\Xhline{1pt}
\rowcolor{gray!16} \textbf{Method} & \textbf{HR@1} & \textbf{HR@2} & \textbf{HR@3} & \textbf{HR@4} & \textbf{HR@5} & \textbf{LCR@1} & \textbf{LCR@2} & \textbf{LCR@3} & \textbf{LCR@4} & \textbf{LCR@5} & \textbf{MRR}\\
\hline
SCASRec (w/o SCR \& EOR) & 71.27 & 87.24 & 88.58 & 93.60 & 95.52 & 82.67 & 89.73 & 90.98 & 93.29 & 93.96 & 0.584 \\
SCASRec (full) & \textbf{71.56} & \textbf{87.78} & \textbf{89.92} & \textbf{95.19} & \textbf{96.98} & \textbf{82.84} & \textbf{91.48} & \textbf{92.54} & \textbf{94.52} & \textbf{94.96} & \textbf{0.590} \\
\Xhline{1pt}
\end{tabular}}
\label{tab:ablation_scr_lr}
\end{table*}

\begin{figure*}[t]
\centering
\includegraphics[width=\textwidth]{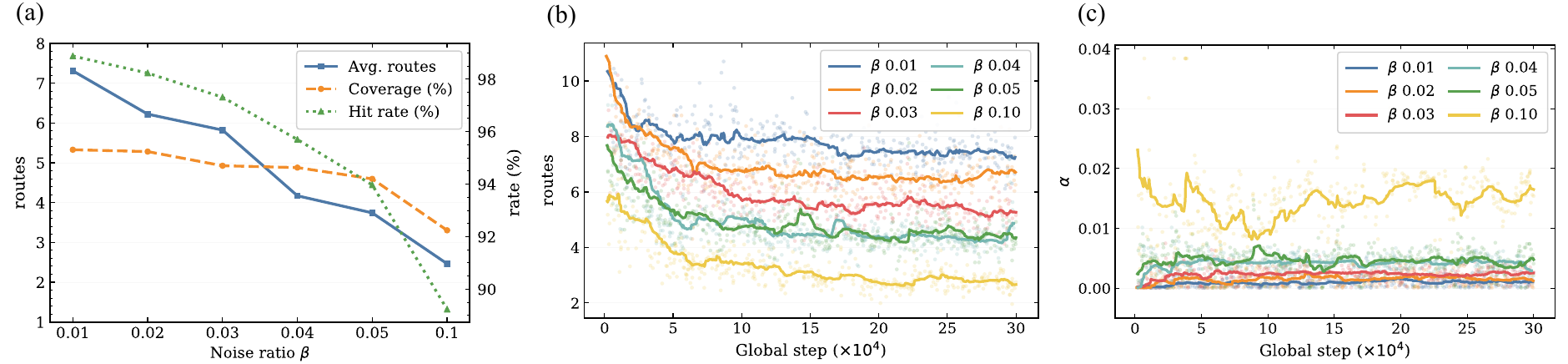}
\vspace{-0.6cm}
\caption{Impact of different overall estimated noise ratio $\beta$ on SCASRec performance.}
\label{fig:hyperparams}
\end{figure*}

Finally, we compare the two training paradigms.
As shown in the last two rows of both tables, the supervised learning (SL) variant consistently outperforms its RL counterpart across nearly all metrics.
For instance, on the MSDR dataset, SCASRec (SL) achieves 56.48\% HR@1, compared to only 41.87\% for SCASRec+RL.
Although reinforcement learning offers a principled framework for directly optimizing non-differentiable list-level objectives, the supervised approach proves significantly more stable and data-efficient in practice, leading to superior convergence and overall performance.
This empirical finding underscores our choice of supervised learning as the core optimization strategy for SCASRec.

\subsection{Ablation Study}
To explore the effectiveness of the core components in SCASRec, we conduct ablation studies on our dataset.
The following analyses examine the impact of core components and hyperparameter settings on model performance.

\subsubsection{Impact of SCR and EOR}
The full SCASRec model leverages SCR to provide stepwise feedback on the marginal gain of improving the current partial list, while the EOR token enables dynamic termination.
To assess their joint contribution, we compare the full model against a variant that disables both mechanisms, effectively reducing SCASRec to a standard autoregressive ranker without list-level correction or adaptive stopping.

As shown in Table~\ref{tab:ablation_scr_lr}, the full model consistently outperforms the ablated version across all metrics.
It achieves higher HR@1 at 71.56\% versus 71.27\% and HR@2 at 87.78\% versus 87.24\%, indicating improved top-rank accuracy.
The full model further dominates at HR@3 through HR@5 with gains of 1.34\%, 1.59\%, and 1.46\% respectively, demonstrating superior list completeness.
More importantly, the full model leads in LCR@K across all $K$, with improvements ranging from 0.17\% at LCR@1 to 1.00\% at LCR@5, showcasing enhanced alignment with user travel trajectories.
The MRR also improves from 0.584 to 0.590, confirming more accurate placement of relevant routes. Furthermore, the case study in Sec.~\ref{sec:case_study} reveals that SCR-guided refinement promotes route diversity.
Since similar routes yield diminishing corrective rewards due to overlapping coverage with the ground-truth trajectory, the model is incentivized to explore meaningfully distinct alternatives offering complementary utility, such as a slightly longer but smoother highway route versus a shorter urban shortcut.
This behavior emerges naturally from the marginal gain formulation of SCR without requiring explicit diversity constraints or post-hoc filtering rules.

In summary, integrating SCR and EOR establishes an effective generative refinement framework that simultaneously improves ranking quality and recommendation diversity, validating the design of SCASRec's self-correcting and auto-stopping mechanism.

\begin{table*}[t]
\centering
\caption{Performance comparison between SCASRec and the online method.}
\vspace{-0.3cm}
\renewcommand\arraystretch{1.1}
\begin{tabular}{
    >{\raggedright\arraybackslash}p{2.1cm}
    *{2}{>{\centering\arraybackslash}p{1.2cm}}
    >{\centering\arraybackslash}p{1.6cm}
    *{4}{>{\centering\arraybackslash}p{1.2cm}}
}
\Xhline{1pt}
\rowcolor{gray!16} \textbf{Method} & \textbf{HR@1} & \textbf{LCR@1} & \textbf{LCR@ALL} & \textbf{Routes} & \textbf{DR} & \textbf{LDR} & \textbf{RRR}\\
\hline
Online Method & 66.67 & 77.56 & 84.50 & 4.313 & 41.81 & 1.231 & 0.211 \\
SCASRec & \textbf{66.75} & \textbf{77.63} & \textbf{85.11} & \textbf{4.171} & \textbf{41.65} & \textbf{0.743} & \textbf{0.104} \\
\Xhline{1pt}
\end{tabular}
\label{tab:online_ab1}
\end{table*}

\subsubsection{Impact of Overall Estimated Noise Ratio $\beta$}
The reward $\alpha$ for EOR in SCASRec is a hyperparameter manually set.
To address this, we designed a noise-aware $\alpha$-adaptation mechanism that dynamically adjusts the model's stopping tendency of the recommendation process.
This mechanism requires a global noise ratio estimate $\beta$, which represents the assumed fraction of noisy (e.g., misclick) samples in the dataset.
Fig.~\ref{fig:hyperparams} summarizes the effect of different $\beta$ values on model performance.

Fig.~\ref{fig:hyperparams}(a) shows the static evaluation across $\beta$.
As $\beta$ increases, the average number of recommended routes decreases, and both coverage and hit rate decline.
This trend is likely caused by the larger reward assigned to the EOR action at higher $\beta$ values, which encourages the model to terminate generation earlier and therefore shortens the recommendation list.

Fig.~\ref{fig:hyperparams}(b) provides a more detailed view of the training dynamics.
First, for any fixed training step, a larger $\beta$ consistently yields a smaller average list length.
Second, across all $\beta$ settings, the average number of recommended routes exhibits a steady downward trend throughout training before eventually converging.
This indicates that the model progressively learns to be more concise as it better discerns user intent and refines its stopping policy.
The convergence point is notably lower for higher $\beta$ values, reinforcing the role of $\beta$ as a control knob for list conciseness.

Fig.~\ref{fig:hyperparams}(c) presents the corresponding $\alpha$ trajectories, where higher $\beta$ produces larger and more volatile learned $\alpha$.
A likely explanation is that an inflated $\beta$ causes the model to assume a higher noise prevalence, which makes it harder to distinguish noisy from informative samples and therefore increases variance in the learned stopping signal, producing less stable learning.
Given that such extreme noise levels are uncommon in real-world applications, we adopt a conservative setting of $\beta=0.04$ in our production deployment.

\subsection{Online Experiments}
We conduct an online A/B test in a widely used navigation application in China to compare SCASRec against the deployed \textit{Online Method}, which integrates PRM, DSFNet, and expert-defined redundancy elimination rules, serving as the strongest baseline in production prior to this work.
As shown in Table~\ref{tab:online_ab1}, SCASRec achieves consistent improvements across all evaluated metrics.
Notably, it reduces the average number of presented routes while simultaneously improving HR@1 and LCR@ALL, indicating better alignment with users’ actual travel behavior.

More importantly, SCASRec significantly enhances recommendation quality from an operational perspective.
It lowers the DR, reduces redundant suggestions, and improves inter-route diversity without relying on any handcrafted rules.
Specifically, SCASRec achieves a 39.6\% reduction in LDR and a 50.7\% reduction in RRR.

These demonstrate that SCASRec not only delivers more accurate route recommendations but also inherently promotes diversity and conciseness through its self-correcting generative design, leading to tangible gains in user experience and system efficiency.

\begin{figure}[t]
\centering
\includegraphics[width=0.47\textwidth]{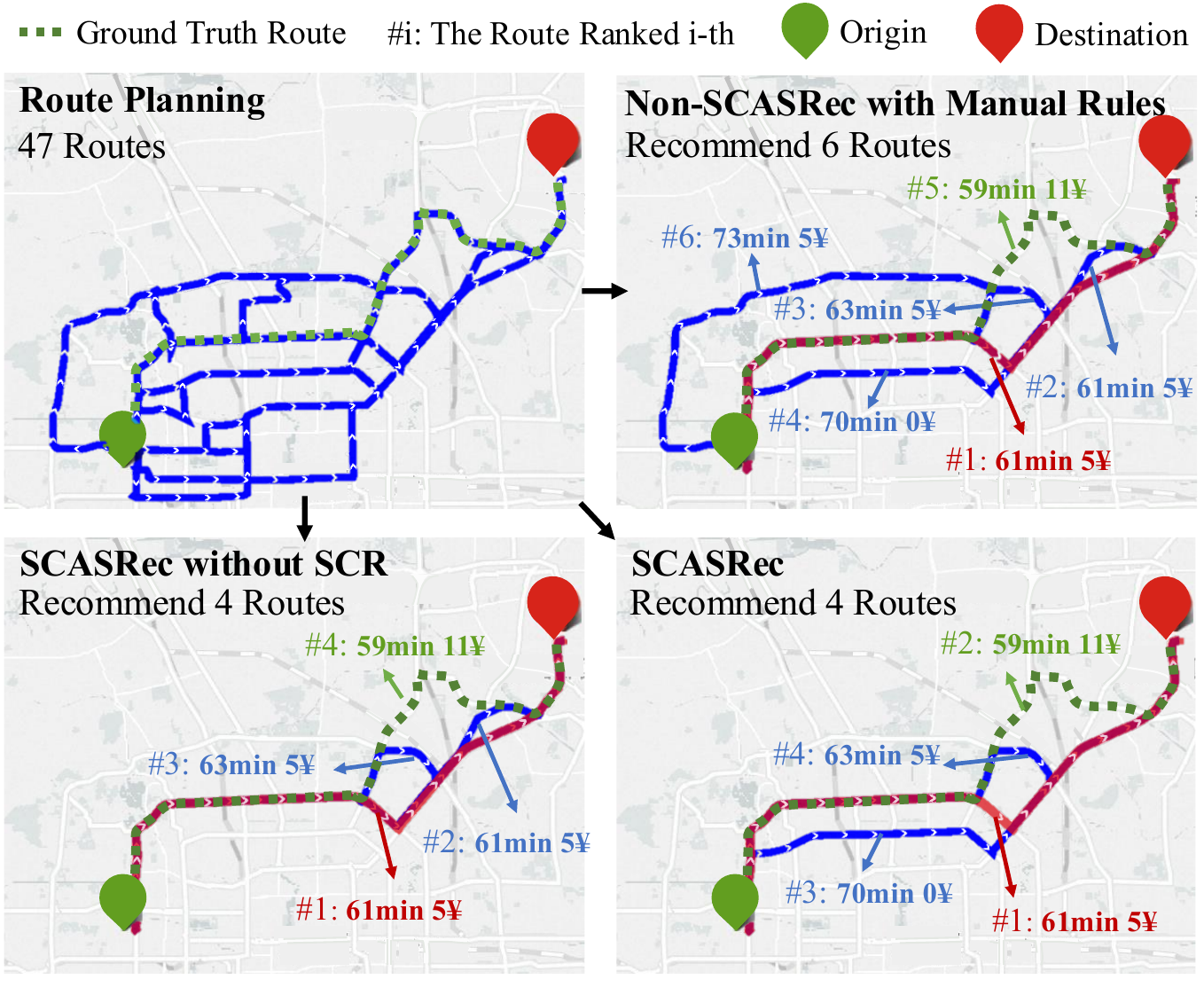}
\vspace{-0.4cm}
\caption{Performance on a real-world recommendation case.}
\vspace{-0.3cm}
\label{fig:case}
\end{figure}

\subsection{Case study}
\label{sec:case_study}
Fig.~\ref{fig:case} presents a challenging real-world route recommendation scenario involving a new user with no historical interaction data.
Given an origin–destination pair, the recall stage retrieves 47 candidate routes.
The ground-truth trajectory is marked by a green dashed line and is difficult to rank due to its suboptimal cost-effectiveness (e.g., longer ETA despite zero toll).
For clarity, we visualize only two key attributes: estimated time of arrival (ETA) and toll cost.

We compare three settings:
(1) Non-SCASRec models: These methods produce top-ranked routes that cluster around a similar trade-off between ETA and toll, resulting in high redundancy.
Although post-hoc expert-defined rules are applied to filter duplicates, they are inflexible and fail to fully resolve similarity (e.g., a redundant route appears at position \#6 even though the ground truth is ranked \#5).
(2) SCASRec without SCR: By leveraging its generative architecture, this variant automatically avoids redundant suggestions without manual intervention.
However, without stepwise corrective feedback, it converges slowly and places the ground-truth route at position \#4.
(3) Full SCASRec: Equipped with SCR, the model rapidly refines its list, ranking the ground truth at \#2.
Moreover, it includes a zero-toll alternative at \#3, demonstrating foresight for potential user rejections of the top candidates.

This case illustrates that SCASRec not only accelerates convergence toward the ground truth but also inherently promotes diversity through its self-correcting mechanism, which eliminates the need for handcrafted redundancy rules and maintains adaptability to complex routing preferences.

\section{Conclusion}
List-wise route recommendation systems are often hindered by three intertwined challenges: the absence of effective list-level supervision, reliance on rigid handcrafted rules for redundancy control, and fragmented optimization across separate ranking stages.
To address these issues in a unified manner, we propose SCASRec, an end-to-end generative framework that jointly performs ranking refinement and redundancy elimination.
By introducing SCR, SCASRec leverages implicit list-wise signals to guide iterative improvement, overcoming the limitations of sparse item-level feedback.
Meanwhile, its learnable EOR token enables adaptive termination without fixed-length assumptions or external heuristics.
Experiments show that SCASRec consistently enhances ranking accuracy and list diversity in both offline and online settings, significantly reducing redundant and low-diversity recommendations.
The model has been successfully deployed in a large-scale production system serving hundreds of millions of daily requests, showing its effectiveness, robustness, and real-world applicability.
Looking ahead, several promising directions emerge.
First, extending SCASRec to multi-modal inputs could further enhance context awareness.
Second, the generative paradigm opens the door to interactive route recommendation.
Third, the core ideas of SCR and EOR are not limited to navigation.
We hope this work inspires more research into unified, generative approaches for list-wise recommendation in real-world applications.

\clearpage
\bibliographystyle{ACM-Reference-Format}
\bibliography{sample-base}

\clearpage
\appendix

\section{Related Works}
\label{app:rwk}
\subsection{Fine-ranking}
Fine-ranking has seen substantial development in recommendation systems, with user sequence modeling and multi-expert models being two of the most active research directions in recent years.
In the field of user sequence modeling, early studies utilized pooling techniques~\cite{covington2016deep} to compress and leverage historical sequence information.
Subsequently, methods based on interest attention mechanisms~\cite{zhou2018deep}, sequential models~\cite{zhou2019deep}, and Transformer-based approaches~\cite{feng2019deep, chen2019behavior, chang2023twin} have further advanced the effectiveness of user sequence modeling.
Recently, inspired by the architectures of LLMs, this field has been gradually shifting towards generative approaches, framing recommendation as a sequence decoding task~\cite{rajput2023recommender, deng2025onerec}.
To tackle the challenges posed by multi-task and multi-scenario recommendation, significant progress has also been made in the realm of multi-expert models.
Examples include the MoE~\cite{jacobs1991adaptive} framework and its various variants~\cite{jacobs1991adaptive, ma2018modeling, tang2020progressive}, as well as more recent models such as STAR~\cite{sheng2021one} for advertising and HiNet~\cite{zhou2023hinet} for e-commerce.

\subsection{Re-ranking}
The most straightforward re-ranking methods balance the diversity of the ranked list by combining relevance scores with manually defined similarity measures, such as DPP~\cite{chen2018fast} and MMR~\cite{carbonell1998use}.
To better capture the contextual information between items, models like SeqSlate~\cite{bello2018seq2slate} based on LSTM~\cite{hochreiter1997long}, DLCM~\cite{ai2018learning} based on GRU~\cite{chung2014empirical}, and PRM~\cite{pei2019personalized} based on Transformer have subsequently been proposed.
However, these methods typically rely on the fine-ranking order as input, leading to an iterative coupling issue.
To better explore and evaluate more permutations, several methods based on generator-evaluator frameworks have been proposed, in which the generator generates multiple permutations, while the evaluator evaluates their quality, such as GRN~\cite{feng2021grn}, PIER~\cite{shi2023pier}, and NAR4Rec~\cite{ren2024non}.
However, these methods are costly as they rely on extensive ground-truth permutations or require training accurate evaluators, and cannot directly optimize for ranking items with higher user interaction probabilities toward the top.

\section{Detailed Problem Formulation}
\label{app:problem_formulation}
This appendix provides the formal definitions of the key metrics used in our problem formulation.

\subsection{Coverage Rate (CR)}
For a recommended $p$ and the user's actual trajectory $u$, the CR~\cite{cui2018personalized, wang2019empowering} is defined as:
\begin{equation}
\text{CR} = \frac{|p \cap u|}{|p \cup u|},
\label{eq:CR}
\end{equation}
which measures the Jaccard similarity between the two paths.
Higher CR indicates that the route better matches users’ expectations.
Let $P^{\text{CR}}=\left\{p^{\text{CR}}_1,\dots,p^{\text{CR}}_N\right\}$ represent the $\text{CR}$ values corresponding to each route in $P$, where the route with the highest CR, denoted as $\hat{p}$ and its CR is denoted as $\hat{p}^{\text{CR}}$, is considered to be the ground truth.

\subsection{Mean Reciprocal Rank (MRR)}
MRR evaluates the ranking position of the ground truth $\hat{p}$.
The closer the rank of $\hat{p}$ is to the top, the better the recommendation performance.
Assuming the dataset is $D$, MRR is calculated as:
\begin{equation}
\text{MRR}(D) 
\!=\! \frac{1}{|D|} \!\sum_{d=1}^{|D|} \text{MRR}\left(\hat{p}_d\right) 
\!=\! \frac{1}{|D|} \!\sum_{d=1}^{|D|} \frac{1}{\text{rank}\left(\hat{p}_d\right)},
\label{eq:MRR}
\end{equation}
where $\hat{p}_d$ denotes the selected route for sample $d$, and $\text{rank}(\hat{p}_d)$ represents the ranking position of $\hat{p}_d$.

\subsection{List Coverage Rate (LCR)}
LCR measures the overall quality of $\bar{P}_d$:
\begin{equation}
\text{LCR}(D) 
\!=\! \frac{1}{|D|} \sum_{d=1}^{|D|} \text{LCR}\left(\bar{P}_d\right) 
\!=\! \frac{1}{|D|} \sum_{d=1}^{|D|} \max_{\bar{p}_i \in \bar{P}_d}\bar{p}^{\text{CR}}_i.
\label{eq:LCR}
\end{equation}
Unlike MRR, LCR can assess list quality even if $\hat{p}$ is not exposed, and it provides a finer-grained evaluation when MRR scores are equal.

\subsection{Redundant Item Exposure}
We define the set of redundant items $Z$ as all routes ranked lower than the ground truth:
\begin{equation}
Z = \bigcup_{d=1}^{|D|} \left\{p_{di}|p_{di}\in\bar{P}_d, \text{rank}\left(p_{di}\right) > \text{rank}\left(\hat{p}_{d}\right)\right\}.
\label{eq:Z}
\end{equation}

\section{Detailed Implementation of SCASRec}
\label{app:detailed_model}

\subsection{Feature Process}
\label{app:feature_process}
The features used in this work include route features, scene features, and user historical sequence features. 
Route features $X^F = \{x^F_1, \dots, x^F_i, \dots, x^F_N\}$ are used to describe each route, including static features, dynamic features, and trajectory statistical features. 
Scene features $E$ represent the contextual information of route recommendation, such as request time, destination POI type, etc.
User historical sequence $H=[h_1,...,h_i,...,h_M]$ is a series of route selection records arranged in chronological order.

As illustrated in the feature processing module in Fig.~\ref{fig:framework}, route features $X^F$ and user historical sequences $ H $ are processed and concatenated to obtain the representation of each route in the candidate set $ X $, denoted as $ X^{en} = \{x_1^{en},...,x_i^{en},...,x_N^{en}\}$.
For simplicity, we still use $ F $ to denote the feature dimension, and $ X^{en}  \in \mathbb{R}^{N \times F}  $ represents the input feature matrix of the candidate set.
Scene features, after processing, are still represented as $E$.
The processing methods for each type of feature are as follows.

\textbf{Route features $X^F$ and Scene features $E$}.
Discrete attributes are processed using embeddings, while continuous attributes are normalized with the z-score method.

\textbf{User historical sequences $H$}.
Each element in $H$ includes the historical scene features and the features of the route selected by the user.
Each $x_i^F\in X^F$, concatenated with $E$, is processed through the DIN~\cite{zhou2018deep} with $H$, resulting in the user historical preference representation $x_i^{h}$.
Finally, $x_i^{h}$ is concatenated with $x_i^F $ to obtain $x_i^{en}$. 

\subsection{Encoder}
\label{sec:encoder}
The encoder takes $ X^{en} $ as input and employs self-attention to capture interactions among candidate routes.
Instead of the feed-forward layer, we incorporate DSFNet~\cite{yu2025dsfnet}, a recently proposed multi-scenario framework designed to generate network parameters with scene features $E$ as input.
After encoding, the output $S^{en}\in\mathbb{R}^{N\times F}$ is the global contextual representation of items and remains constant throughout a single forward generation process.
$S^{en}$ will serve as part of the state representation in the decoder. 

\subsection{Decoder}
\label{sec:decoder}
The decoder first appends a virtual route EOR to $X^{en}$, representing the stopping signal of the generation process.
The representation of EOR is initialized as a learnable vector of length $F$. We use $ X^{de}\in\mathbb{R}^{(N+1)\times F}$ to denote the candidate routes set in the decoder. 
The decoder iteratively generates selection probabilities for the remaining candidate items to construct the recommendation list.
Suppose at the step $t,t\geq 1$, the list of items already generated is $\bar{P}_t$.
The computation process for the selection probabilities $P_t$ of candidate routes set $X^{de}$ at step $t$ is as follows.

Firstly, through a look-up layer, the features of routes in $\bar{P}_t$ are retrieved from $X^{de}$ and extracted as the representation $\bar{X}_t^{de}\in\mathbb{R}^{t\times F}$.
When $t=1$, $\bar{P}_t$ is empty, a learnable vector of length $F$ is initialized and added to represent the start.

Secondly, to capture the contextual relationship between candidate routes $X^{de}$ and $\bar{X}_t^{de}$, we designed a state attention mechanism.
State attention treats each $x_i^{de}\in X^{de}$ as $Q$, and $\bar{X}_t^{de}$ as $K$ and $V$, to derive the representation of the stepwise contextual relationship $S^{de}_t\in\mathbb{R}^{(N+1)\times F}$:
\begin{equation}
S^{de}_t = \sigma\left( X^{de} W^Q \left(\bar{X}_t^{de} W^K\right)^\top\right) \bar{X}_t^{de} W^V,
\label{eq:sde}
\end{equation}
where $ W $ represents the linear transformation parameters.
The sigmoid function $\sigma$ is used as the attention function instead of softmax because the attention values need to account for both the items in $\bar{X}_t^{de}$ and the size of $\bar{X}_t^{de}$, and therefore should not be restricted to sum to 1.

Thirdly, we add EOR into $S^{en}$, and then concatenate it with $S^{de}_t$ to obtain the state representation $S_t\in\mathbb{R}^{(N+1)\times F}$.
Finally, $P_t$ is calculated after DSFNet and softmax:
\begin{equation}
\begin{aligned}
&\quad\quad\quad\quad S^{en}=\text{concat}\left(S^{en},\text{EOR}\right), \\
&\quad\quad\quad\quad\;\; S_t=\text{concat}\left(S^{en}, S^{de}_t\right), \\
&P_t=\text{softmax}\left(\text{DSFNet}\left(S_t, E\right)+\text{mask}_{\bar{P}_t}\right),
\end{aligned}
\label{eq:pt}
\end{equation}
where the purpose of the mask is to add $-\infty$ to items recommended in $\bar{P}_t$, preventing duplicate recommendations.

\subsection{Noise-aware $\alpha$-adaptation}
\label{appendix:alpha}
The reward $\alpha$ for EOR is a hyperparameter that controls the stopping tendency of the recommendation process.
A higher $\alpha$ results in earlier stops and fewer recommended items.
Since $\alpha$ lacks a physical interpretation, it is difficult to set it directly.
Therefore, we design a noise-aware $\alpha$-adaptation algorithm.
The ``noise" refers to unpredictable noise (e.g., user misclicks), which is difficult to remove manually from the dataset.
Due to the incorrect labels, these noisy samples are typically harder for the model to learn.

Assume that under a given $\alpha$, the dataset $D$ is divided into two sets, $D_{suc}$ and $D_{fail}$, representing the samples for which the $\hat{x}$ are successfully and unsuccessfully recommended, respectively.
Generally, samples in $D_{fail}$ are harder to learn and are more likely to contain noise.
The proportion of $D_{fail}$ is defined as $e=\frac{|D_{fail}|}{|D|}$, and the overall estimated noise ratio in the dataset is denoted as $\beta$. 

The noise-aware $\alpha$-adaptation is a heuristic algorithm that adapts $\alpha$ dynamically during training, aiming to make $e$ approximate $\beta$.
Specifically, after each batch of training, we compute $e$ and update $\alpha$ based on the following formula:
\begin{equation}
\alpha = 
\begin{cases} 
\alpha + \eta, & \text{if }  e < \beta , \\
\max (\alpha - \eta, 0), & \text{if }  e > \beta , \\
\alpha , & \text{if }  e = \beta .
\end{cases}
\label{eq:alpha}
\end{equation}
Since the heuristic algorithm converges quickly, $\eta$ can simply be assigned a small value.
In our work, $\eta=1e\text{-}4$.
$\beta$ has a clear physical interpretation, representing the estimated noise ratio, and can be set based on specific scenarios.

\begin{algorithm}[t]
\caption{The training process of SCASRec}
\label{alg}
\begin{algorithmic}
\REQUIRE All epochs, parameters $\theta$, maximum steps $T$
\ENSURE Optimized $\theta$
\FOR{each mini-batch $D \subseteq \text{epochs}$}
    \STATE Get $X^{\text{en}}, E$ via feature processing
    \STATE $S^{\text{en}} \gets \text{Encoder}\left(X^{\text{en}}\right)$
    \STATE $X^{\text{de}} \gets \text{concat}\left(X^{\text{en}}, \text{EOR}\right)$
    \STATE $S^{\text{en}} \gets \text{concat}\left(S^{\text{en}}, \text{EOR}\right)$
    \STATE Initialize $D_{\text{fail}}\gets\emptyset$, $\mathcal{L}\gets 0$
    \STATE Initialize $\bar{P}\gets\left[\left[\text{start}\right]\times |D|\right]$, $\hat{t}\gets\left[\left[T+1\right]\times |D|\right]$
    \FOR{$t=1$ to $T$}
        \STATE Get $P_t, Y_t, r_t$ via Eq.~\eqref{eq:sde},~\eqref{eq:pt},~\eqref{eq:scr}-\eqref{eq:loss}
        \STATE $\mathcal{L}\gets\mathcal{L}-r_t \cdot Y_t \cdot\log (P_t)$
        \STATE Remove samples where $t=\hat{t}+1$
        \IF{no samples left}
            \STATE break
        \ENDIF
        \STATE Select the non-EOR item with the largest value in $P_t$ and append to $\bar{P}$
        \FOR{each $d \in D$}
            \IF{$\hat{p}$ in $\bar{P}$}
                \STATE $\hat{t} \gets t$
            \ENDIF
            \IF{$\arg\max P_t=\text{EOR}$}
                \STATE Add $d$ to $D_{\text{fail}}$
            \ENDIF
        \ENDFOR
    \ENDFOR
    \STATE $\theta \gets \text{Adam}(\theta,\mathcal{L})$
    \STATE Update $\alpha$ via Eq.~\eqref{eq:alpha}
\ENDFOR
\end{algorithmic}
\end{algorithm}

\subsection{Reinforcement Learning Variant}
\label{app:rl}
For comparative purposes, we also implement a reinforcement learning (RL) variant of SCASRec, denoted as SCASRec+RL.
In this variant, the model is trained to directly maximize the cumulative reward derived from the list-level objective.

Specifically, during training, actions (routes or the EOR token) are stochastically sampled from the model's output distribution, enabling exploration.
The generation terminates upon sampling the EOR token.
The stepwise reward is defined as:
\begin{equation}
r_t = 
\begin{cases}
r_t^{\text{SCR}}, & \text{if } t \leq \hat{t}, \\
-\alpha,          & \text{if } t > \hat{t} \text{ and } p_t \neq \text{EOR},
\end{cases}
\label{eq:rl_reward}
\end{equation}
where $\hat{t}$ is the step at which the ground-truth item $\hat{p}$ is recommended, $r_t^{\text{SCR}}$ is as defined in Eq.~\eqref{eq:scr}, and $\alpha$ penalizes redundant recommendations after the ground truth has been recommended.

To align with the nature of MRR, which favors earlier positions, we employ a discounted return with $\lambda=0.5$:
\begin{equation}
Q_t = \sum_{k=t}^{T} \lambda^{k-t} r_k,
\label{eq:discounted_return}
\end{equation}
where $T$ denotes the step at which \text{EOR} is sampled.
Since MRR assigns higher scores to earlier positions of the ground-truth item, the discounted return naturally emphasizes earlier recommendations when $\lambda < 1$.
By setting $\lambda = 0.5$, we strengthen the preference for early correction, thereby directly encouraging the model to improve MRR through temporal credit assignment.

We employ the REINFORCE algorithm~\cite{williams1992simple} to update the policy parameters $\theta$, maximizing the expected cumulative reward.
The gradient is given by:
\begin{equation}
\nabla_\theta \mathcal{J}_{\text{RL}}(\theta)=\mathbb{E}_{\bar{X}\sim\pi_\theta}
\left[\sum_{t=1}^{T} \nabla_\theta \log \pi_\theta(p_t | s_t) \cdot Q_t \right].
\label{eq:reinforce}
\end{equation}
This RL variant serves as a baseline to highlight the advantages of our primary supervised learning approach.

\section{Extended Theoretical Discussion}
\label{app:theory}
A conventional two-stage pipeline is structurally incapable of reliably recovering $\pi^*$.
The fine-ranking stage learns an item-wise scoring function $s: \mathcal{P} \to \mathbb{R}$ in isolation, without knowledge of the final list-level objective $F$.
If there exists any candidate route $p' \in \mathcal{P}$ such that $s(p')>s(\hat p)$, which is highly probable if $s(\cdot)$ is trained on proxies that may not perfectly correlate with $\text{CR}(\cdot)$ or the true user intent for the current session, then $\hat p$ will not be ranked at the top by the fine-ranker.
The subsequent re-ranking stage operates only on a fixed, truncated set of candidates from the fine-ranker's output.
Even if the re-ranker is aware of the list-level goal, its ability to promote $\hat p$ to the first position is fundamentally constrained by the initial ranking and the pre-defined list length.
Moreover, the pipeline lacks an integrated mechanism for adaptive termination, making it impossible to dynamically achieve $|Z| = 0$ in a data-driven manner.
Consequently, the two-stage approach is confined to a subspace of solutions that are locally optimal with respect to its decomposed stages but globally sub-optimal with respect to $F$.

\section{Datasets}
\label{app: dataset}
We evaluate SCASRec on two real-world route recommendation datasets: 
(1) a large-scale proprietary dataset collected from a major Chinese navigation platform, which we introduce and will publicly release; and 
(2) the open-source MSDR (Multi-Scenario Driving Recommendation) benchmark, recently proposed by AMap.
Both datasets provide rich contextual features, user interaction logs, and trajectory-derived ground-truth labels, enabling rigorous offline and online evaluation of list-wise recommendation models.

The dataset was collected from an online navigation application in China and comprises approximately 427 thousand users, 512 thousand samples, and 6 million routes across 370 cities.
It includes user information, recalled route information, user preferences, user familiarity with the recalled routes, and navigation scenario details.
Table~\ref{tab:datafeat} contains the feature dimensions and key features of our dataset.
Additionally, the dataset contains users' actual travel data, which is reflected in the coverage rate between the recalled routes.
This route recommendation dataset can be utilized for fine-ranking and re-ranking tasks, providing valuable data support for developing more accurate and effective route recommendation systems.
Our dataset will be publicly available on Google Drive \footnote{\url{https://drive.google.com/drive/folders/1Ku3DE2YmHgrpskpgU6PpuzBhQ_2qwzPg}}.

The MSDR dataset was constructed from two weeks of driving navigation logs (June 25 – July 8, 2023) collected by AMap across eight major Chinese cities (i.e., Beijing, Shanghai, Guangzhou, Hangzhou, Wuhan, Zhengzhou, Chongqing, Chengdu).
For each navigation session, up to 100 candidate routes are recalled, and the top three are presented to the user.
The selected route and off-route deviation points are used to reconstruct the ground-truth trajectory.
The route with the highest coverage rate against this trajectory is labeled positive, while a balanced set of negative samples is retained.
MSDR provides rich features, including route geometry, real-time traffic conditions, POI categories, user demographics, and scenario context (e.g., time, congestion, origin/destination types), making it a valuable public benchmark for route recommendation research.

\end{document}